\documentclass[10pt,twocolumn]{IEEEtran}
\makeatletter
\def\ps@headings{%
\def\@oddhead{\mbox{}\scriptsize\rightmark \hfil \thepage}%
\def\@evenhead{\scriptsize\thepage \hfil \leftmark\mbox{}}%
\def\@oddfoot{}%
\def\@evenfoot{}}
\makeatother 

\usepackage{graphicx,amssymb,amsmath,amsfonts}
\usepackage{subfigure}
\begin{document}

% the first alternative ...

\title{Joint Cooperation and Multi-Hopping Increase the Capacity of Wireless Networks}
\author{Sam Vakil,~\IEEEmembership{Student Member,~IEEE}, and Ben Liang,~\IEEEmembership{Senior
Member,~IEEE}
\thanks{The authors are affiliated with the Department of Electrical and Computer Engineering,
University of Toronto, 10 King's College Road, Toronto, Ontario, M5S 3G4, Canada.
Email: \{vakil,liang\}@comm.utoronto.ca.
This work was supported in part by a grant
from LG Electronics.  A preliminary version of this paper has appeared in
the IEEE Communications Society Conference on Sensor, Mesh and Ad Hoc Communications and Networks (SECON) 2008.
This version contains additional details and expanded analysis on different network environments.
} }

\date{}
\maketitle \thispagestyle{empty}

\begin{abstract}
The problem of communication among nodes in an \emph{extended
network} is considered, where radio power decay and interference
are limiting factors.  It has been shown previously that, with
simple multi-hopping, the achievable total communication rate in
such a network is at most $\Theta(\sqrt{N})$. In this work, we
study the benefit of node cooperation in conjunction with
multi-hopping on the network capacity. We propose a multi-phase
communication scheme, combining distributed MIMO transmission with
multi-hop forwarding among clusters of nodes.  We derive the
network throughput of this communication scheme and determine the
optimal cluster size. This provides a constructive lower bound on
the network capacity. We first show that in \textit{regular
networks} a rate of $\omega(N^{\frac{2}{3} })$ can be achieved
with transmission power scaling of
$\Theta(N^{\frac{\alpha}{6}-\frac{1}{3}})$, where $\alpha>2$ is
the signal path-loss exponent. We further extend this result to
\textit{random networks}, where we show a rate of $\omega
\big(N^{\frac{2}{3}}(\log{N})^{\frac{2-\alpha}{6}}\big)$ can be
achieved with transmission power scaling of
$\Theta(N^{\frac{\alpha}{6}-\frac{1}{3}}(\log{N})^{-\frac{(\alpha-2)^2}{6}})$
in a random network with unit node density. In particular, as
$\alpha$ approaches 2, only constant transmission power is
required. Finally, we study a random network with density
$\lambda=\Omega(\log{N})$ and show that a rate of $\omega((\lambda
N)^{\frac{2}{3}})$ is achieved and the required power scales as
$\Theta(\frac{N^{\frac{\alpha}{6}-\frac{1}{3}}}{\lambda^{\frac{\alpha}{3}-\frac{2}{3}}})$.
\end{abstract}

\section{Introduction}
Understanding the feasibility of multi-hop wireless networks has
been the subject of a great deal of research. In this regard, much
mathematical and practical consideration has been devoted to
studying the capacity of such networks. In the seminal work of Gupta
and Kumar \cite{kumar}, it has been shown that under the assumption
of point-to-point communication, the per-node asymptotic capacity
decays at least as fast $\frac{1}{\sqrt{N}}$ for a \textit{dense
network} with $N$ nodes, where the number of nodes approaches
infinity within a unit disk or unit sphere.  Subsequent studies have
shown similar decay for upper-bound on capacity in an \emph{extended
network}, where the deployment area increases linearly with the
number of nodes while the node density remains constant
\cite{pramod}\cite{leveq}\cite{xie}. In such networks, besides
interference, the radio power decay over large communication
distances is another determining factor affecting the communication
rate.

The results of \cite{kumar} are based on the assumption that the
communication between a specific source-destination pair is always
deteriorated by the concurrent transmission of the other nodes. In
contrast, cooperative communication can result in considerable
capacity gains by discarding this constraint. Under this setting,
the notion of a link is considered as a set of users, encoding and
transmitting messages in coordination. Such a communication setting
mimics a Multiple-Input Multiple-Output (MIMO) antenna system
\cite{tse2} in a distributed fashion, where users act as
\textit{virtual} antenna arrays cooperating to transmit a message
towards the destination.

Exploiting the spatial multiplexing gain of MIMO is appealing, since
MIMO communication can result in linear capacity increase as a
function of the number of antennas in the high SNR regime
\cite{tse2}.  Furthermore, Ozgur \emph{et al} has recently extended
this result to \textit{distributed} MIMO communication \cite{ozgur2}
to achieve linear capacity scaling with $N$ in a \emph{dense}
network with constant transmission power, by utilizing a
hierarchical communication scheme with single-hop distributed MIMO
transmission between the source and destination clusters. However,
multi-hop forwarding is essential to the efficiency of large-scale
\textit{extended} networks.  It remains an open problem to quantify
the capacity of multi-hop cooperative communication.

This work represents a step toward this direction.  We extend our
results in~\cite{vakil08} and study the benefit of cooperation in
an extended network, by combining peer-node cooperation with
multi-hop forwarding to mitigate the path-loss power decay and
harness spatial multiplexing gain. We propose a multi-phase
communication scheme based on \textit{multi-hop} distributed MIMO
forwarding among clusters of nodes, where a cluster is a
hierarchically organized set of nodes that cooperate
simultaneously to transmit a single message vector. We then derive
the network throughput of this communication scheme by considering
the optimal cluster size. We first show that, for a
\textit{regular network}, where one node is randomly placed within
each square of unit area, the network throughput scales as
$\omega(N^\frac{2}{3})$ with transmission power requirement of
$\Theta(N^{\frac{\alpha}{6}-\frac{1}{3} })$, where $\alpha>2$ is
the signal path loss exponent. We further extend this result to a
\textit{random network}, where the nodes are uniformly distributed
with unit node density, and show that throughput scaling is
lower-bounded by
$\omega\big(N^{\frac{2}{3}}(\log{N})^{\frac{2-\alpha}{6}}\big)$
with transmission power requirement of
$\Theta(N^{\frac{\alpha}{6}-\frac{1}{3}}(\log{N})^{-\frac{(\alpha-2)^2}{6}})$.
We finally study a random network with increasing node density and
show that if the density $\lambda$ follows
$\lambda=\Omega(\log{N})$ the throughput scaling is lower-bounded
by $\omega((\lambda N)^{\frac{2}{3}})$ with power scaling
$\Theta(\frac{N^{\frac{\alpha}{6}-\frac{1}{3}}}{\lambda^{\frac{\alpha}{3}-\frac{2}{3}}})$.
The existence of the proposed communication scheme provides a
constructive lower bound on the capacity of extended networks with
node cooperation.

The rest of the paper is organized as follows. In Section
\ref{related}, we summarize the related work. Section \ref{model}
details the network model. In Section \ref{outline} we explain the
three phases of the communication scheme. Section \ref{cresult}
analytically evaluates each communication phase and presents the
overall network capacity. We extend our results to random networks
in Section \ref{numsim}. Finally, the concluding remarks are given
in Section \ref{conc}.

\section{Related Work}
\label{related}

Existing research on the scaling behavior of multi-hop wireless
networks can be categorized mainly into two groups. The first
category is concentrated on information-theoretic upper-bounds on
the network capacity by using cut-set bounds, and the second
category deals with constructive communication schemes which achieve
a lower-bound on the capacity. Different topologies such as dense
and extended networks have been considered in each category.

\textit{Information Theoretic Upper-bounds}: The seminal work
\cite{kumar} provides both upper-bounds and constructive
communication schemes to show that the point-to-point capacity of a
dense wireless network is $\Theta(\sqrt{N})$. In \cite{pramod},
\cite{leveq}, \cite{xie}, and \cite{xie2}, the authors derive
information theoretic upper-bounds on the network capacity.  These
works study the capacity scaling of extended networks and obtain
upper-bounds on the capacity by using cut-set bounds. In particular,
Xie \emph{et al} prove that for environments with path-loss exponent
$\alpha>4$ and constant per node power, the expected transport
capacity grows at most linearly in the number of nodes, so that for
an extended network with uniformly distributed source-destination
pairs, the network capacity is $O(\sqrt{N})$\cite{xie}.

While in simple multi-hopping, increasing the transmission power
does not improve the capacity scaling, Jovicic \emph{et al} in
\cite{pramod} show that the upper-bound on the network transport
capacity scales linearly with node power. Similar results are shown
in \cite{xie}. Their results motivates using the extra degree of
freedom provided by the choice of transmission power.   We exploit
this in the proposed collaborative scheme to increase the network
capacity.

\textit{Constructive Communication Schemes:} Toumpis \emph{et al} in
\cite{toumpis} have proposed a gridding of a unit area network and
the use of 9-TDMA scheduling to derive the same lower-bound on
capacity as \cite{kumar} in a more straightforward manner.  They
further extend their model to consider the effect of node mobility
and fading on the capacity. Franceschetti \emph{et al} in
\cite{massimo} have closed the previous gap between the capacity in
\emph{arbitrary} and \emph{random} networks. They use percolation
theory to show that the sum capacity of their proposed communication
scheme scales as $\Theta(\sqrt{N})$. None of the above utilizes
cooperative transmission.

There has been much research into node cooperation in the context of
single-source, single-destination, and $N$ relays. For example,
Gastpar \emph{et al} in \cite{gastpar} show an achievable rate
scaling of $\Theta(\log{N})$.  However, Ozgur \emph{et al} in
\cite{ozgur2} are the first to exploit the linear-scaling result of
MIMO communication in a multiple-source network using a distributed
MIMO paradigm. For a \emph{dense} network, they propose a
hierarchical cooperative communication scheme with single-hop
distributed MIMO transmission to achieve a network throughput
scaling of $\Omega(N^{1-\epsilon})$, $\forall \epsilon>0$, and
single-hop distributed MIMO transmission. They further show that
when the same scheme is applied to the more realistic extended
network model, it results in a throughput
$\Theta(N^{2-\frac{\alpha}{2}+\epsilon})$, implying that for
$\alpha<3$, the scheme outperforms simple multi-hopping.

Our work is motivated by \cite{ozgur2}.  However, we propose a
multi-hop distributed MIMO scheme specifically designed for extended
networks.  Furthermore, we show that with a small increase in the
transmission power, our scheme performs uniformly better than simple
multi-hopping for all values of $\alpha>2$.  As far as we are aware,
there is no existing work on analyzing the achievable rate of
combining cooperative communication and multi-hop forwarding.

Finally, Aeron and Saligrama in \cite{aeron} have studied the effect
of node cooperation on the capacity of wireless networks with a
fixed receiver SNR at all nodes and obtained
$\Theta(N^{\frac{2}{3}})$ network throughput, through spatially
separated MIMO relays which collaborate in their transmissions.
However, the assumption of constant receiver SNR at all nodes
implies that the transmission power needs to scale as
$\Theta(N^{\frac{\alpha}{2}})$. In contrast, in this work we show
that with much lower power scaling, the same communication rate can
be achieved by using multi-hop distributed MIMO transmissions.

\section{Network Model}
\label{model} We consider an extended network with $N$ nodes
distributed within the area $B_N=[0,\sqrt{N}]\times[0,\sqrt{N}]$.
We first evaluate the capacity results for nodes following a
special topology that we call a \emph{regular network}. In a
regular network, we divide $B_N$ to $N$ squares of unit area and
assume there is exactly one node in each such square. We then
generalize our results to a \textit{random network}, where nodes
are distributed independently and uniformly over the area $B_N$.

A matching of the source-destination pairs is picked at random, so
that each node is the destination of exactly one source. It is
assumed that the sources are all sending their messages with a
common source-to-destination rate $R(N)$ bits/sec, and the total
network throughput is $T(N)=NR(N)$ bits/sec. Each node divides its
messages into sub-blocks of length $L$ bits and sends packets with
length equal to a multiple of $L$ during the communication with its
intended receiver. It will be clear later that the proper sub-block
size $L$ depends only on the MIMO transmission details and does not
affect the throughput scaling analysis.  In contrast, the multiplier
to $L$ is an optimization parameter dependent on the optimal size of
a cluster which will be explained in detail in
Section~\ref{cresult}.

Similar to the model of
\cite{kumar}\cite{ozgur2}\cite{massimo}\cite{aeron}, we
consider a line-of-sight environment without fading or
shadowing. The channel gain between two nodes $i$ and $j$ in
the network is assumed to follow a standard baseband model and
can be written as
\begin{eqnarray}
\label{channelg}
h_{ij}=\frac{e^{\mathfrak{i}\theta_{ij}}}{d_{ij}^{\frac{\alpha}{2}}},
\end{eqnarray}
where $d_{ij}$ is the Euclidean distance between the two nodes,
$\mathfrak{i}$ is the unit imaginary number, $\theta_{ij}$ is
the phase change  distributed uniformly and independently in
$[0,2\pi]$ \cite{tse2}, and $\alpha> 2$ is the path loss
exponent, such that the power of the received signal at node
$j$ from node $i$ equals $\frac{P_i}{d_{ij}^\alpha}$, where
$P_i$ is the transmission power of node $i$. Then, the received
message at node $j$ at time $m$ can be expressed as
\begin{equation}
\label{modelformul}
\begin{split}
 Y_j[m]=\sum_{i \in
\mathbb{T}[m]}h_{ij}[m]X_{i}[m]+Z_j[m]+I_j[m],
\end{split}
\end{equation}
where $\mathbb{T}[m]$ represents the set of active senders at time
$m$ transmitting signals $X_{i}[m]$ to node $j$, which can be added
constructively \cite{tse2}, $Y_j[m]$ is the received message at node
$j$ at time $m$, $Z_j[m]$ is the additive white Gaussian noise at
node $j$ with variance $N_0$, and $I_j[m]$ is the interference from
the nodes which are destructive to the reception of node $j$.
 We do not consider sophisticated multi-user detection
techniques at the receiver nodes and simply treat interference as
noise to obtain a lower-bound on the network capacity. Furthermore,
since we are proposing a constructive lower-bound on the capacity,
during each communication phase, a network topology which results in
the lowest communication rate is considered. Therefore, the
dependency of the channel gains on time can be removed in our
analysis.

\section{Communication Scheme}
\label{outline} 
\begin{figure}
\vspace{-2mm}
  \centering
\subfigure[]{\includegraphics[scale=.55]{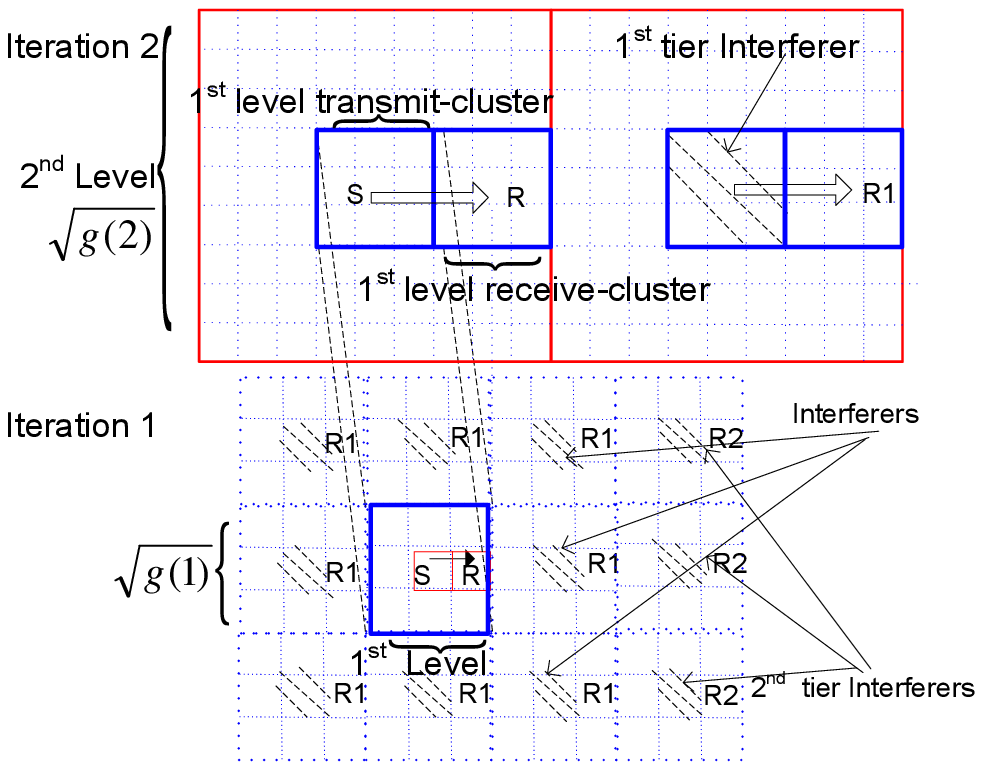}}
\subfigure[]{\includegraphics[scale=.55]{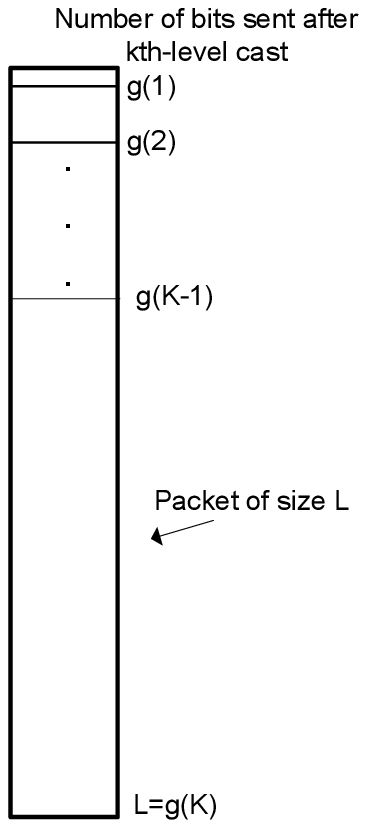}}

  \caption{(a) $k$th-level clusters and pattern of interferers to a specific cluster in the intra-cluster phase. (b) Iterative progression of intra-cluster packet transmission.}\label{intra1}
\vspace{-3mm}
 \end{figure}

We consider a communication scheme whose rate
provides a constructive lower-bound on the network capacity. We
arrange the nodes into a hierarchical $K$ level clustering
structure, each level consisting of 9 non-overlapping lower-level
clusters in a 3-by-3 formation, such that the $k$\emph{th-level
cluster} refers to the set of nodes that lie in a square of area
$g(k)=3^{2k}$, as depicted in Fig.~\ref{intra1}(a). Unless otherwise
stated, we use the term \textit{cluster} synonymously with the term
\textit{$K$th-level cluster}.

The proposed scheme is comprised of three \textit{phases},
intra-cluster message dissemination from the source to all
nodes in its cluster, inter-cluster message forwarding toward
the destination node's cluster, and message decoding at the
destination.  In the following we explain the transmission
scheme in each phase for the regular network. The
generalization to the case of a random network follows the same
structure and the difference is explained in
Section~\ref{numsim}. Throughout the paper, we use the notation
$P^{(i)}$, for the transmission power during the $i$th phase.

 \textbf{Phase 1: Intra-Cluster
Communication}. The goal of this phase is to distribute each
source node's message to the set of $g(K)$ nodes in the
source's cluster (which will serve as relays for the source in
the next phase). We utilize a hierarchical cooperation strategy
carried out in $K$ iterations as shown in Fig.~\ref{intra1}(a).
Each source node divides its message to packets of length
$g(K)L$ bits, (i.e., g(K) sub-blocks and $L$ bits per
sub-block). The packet's transmission is carried out in a
$K$-level progression as shown in Fig.~\ref{intra1}(b), such
that after the $k$th iteration, the source $S$ will have
forwarded $g(k)L$ bits of its packet. We assume that the nodes
transmit with a common constant power $P^{(1)}=P$ during this
phase. The details of this iterative communication scheme are
given in the following.

\begin{figure}
\vspace{-2mm} \centering
\subfigure[]{\includegraphics[scale=.6]{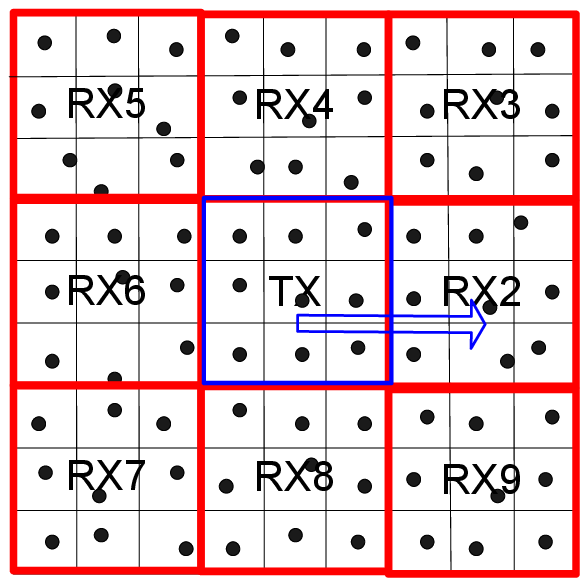}}
\subfigure[]{\includegraphics[scale=.5]{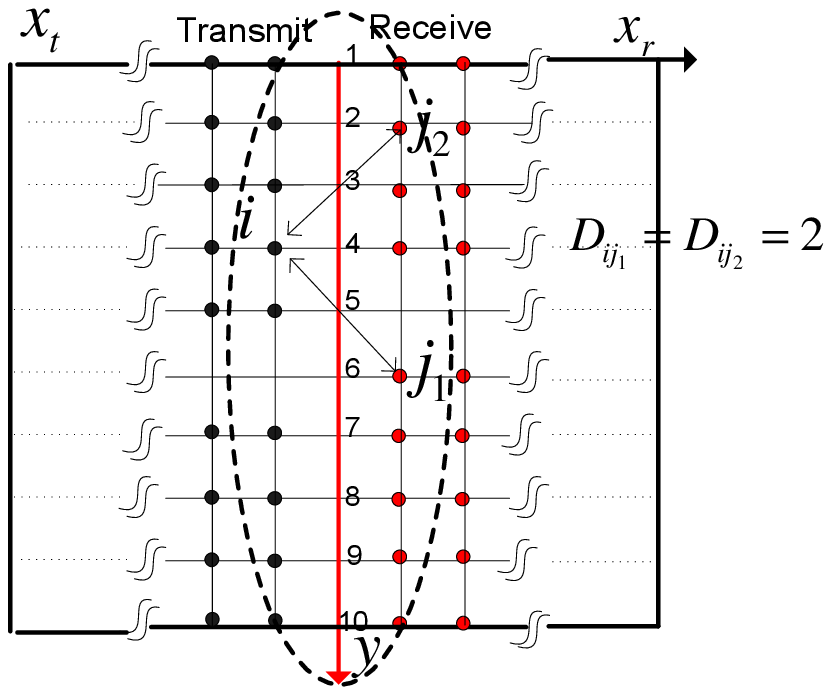}}
\caption{(a) Pattern of transmissions between the neighbor clusters.
(b) Node topology for constructing capacity lower-bound. Center red
line represents border between two 2nd-level clusters} \label{sched}
\vspace{-3mm}
\end{figure}

Two $k$th-level clusters are said to be \emph{neighbors} if their
boundaries touch. Therefore, each $k$th-level cluster has 8
neighbors. To avoid interference from neighboring clusters, TDMA is
used to schedule transmission among the 9 partitioning
$(k-1)$th-level clusters within a $k$th-level cluster.  This is
called a 9-TDMA scheme. As shown in Fig.~\ref{sched}(a), 9-TDMA
implies that, during the $k$th iteration, only one among the nine
$(k-1)$th level clusters acts as the \emph{transmit-cluster}, and
its eight neighbor clusters are potential \emph{receive-clusters}.
The $(k-1)$th level transmit- and receive-clusters communicate using
distributed MIMO transmission. We call each such communication a
\emph{$k$th-level cast}. At the receive-cluster side, during the
$k$th-level cast, each of the $(k-1)$th-level receive-clusters is
set to active receive mode in a round-robbin manner, based on a
specific order as depicted in Fig.~\ref{sched}(a).

During the first iteration, an active source performs 8 consecutive
$1$st-level casts to transmit a different sub-block of $L$ bits to
each of its 8 neighbor nodes and keeps $L$ bits in its own buffer.
Hence, at the end of the first iteration,  the 1st-level cluster
contains $9L$ bits. During the second iteration, as depicted in the
2nd-level in Fig.~\ref{intra1}(a), the $g(1)=9$ nodes inside the
first level transmit-clusters act as distributed antennas, sending
their messages at the same time. Thus, during each $2$nd-level cast,
$9L$ bits are sent simultaneously using distributed MIMO to one of
the 8 neighbor receive-clusters of size $g(1)$. Note that
concurrently with this, the source can distribute another set of
$9L$ bits to its 1st-level cluster in preparation for the next
2nd-level cast.\footnote{This operation assumes the full-duplex
communication, which is common in multi-hop capacity analysis
\cite{kumar}. If only half-duplex communication is available, the
1st-level source-cluster will simply wait for the source to send
sufficient data before each 2nd-level cast. The same procedure can
be applied in all levels, and it is easy to show that the resultant
increase in transmission time does not change the scaling of the
optimal cluster size or the network throughput.} Hence, through 8
consecutive 2nd-level casts, $81L$ bits of the source are
transmitted in this iteration to the 2nd-level cluster. This pattern
repeats, so that in the $K$th iteration, $g(K)L$ bits of the source
are transmitted to the $K$th-level cluster. The number of
iterations, $K$, and therefore the required time, $T_1$, to finish
this step are design parameters to be found in
Section~\ref{cresult}.

Later, to determine the effect of interference on the network
throughput, we note that the set of interferer clusters imposed by
9-TDMA follows the pattern depicted in Fig.~\ref{intra1}(a), where R
represents the $k$th-level receive-cluster. The shaded
transmit-clusters, with corresponding receive-clusters labeled `R1',
act as interferers (1st \emph{tier} interferers), since they are
transmitting synchronously during R's reception. It is clear that in
the $i$th tier there are $8i$ interferers.  (Some of the 2nd tier
interferers are depicted in the figure). As the clusters increase in
size, we still have the same pattern for the interferers' locations.
This can be better seen in Fig.~\ref{intra1}(a) by considering the
2nd-level cast, where the spatial separation of interferers is
multiplied by 3 from that of the 1st-level cast.

\begin{figure}
\vspace{-2mm} \centering
\subfigure{\includegraphics[scale=.6]{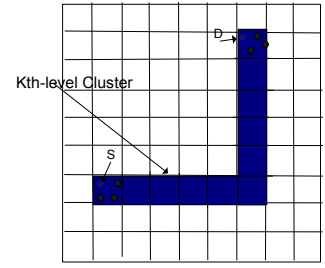}} \caption{
Routing in the inter-cluster phase.} \label{routingg}\vspace{-4mm}
\end{figure}

\textbf{Phase 2: Inter-Cluster Communication}. The goal of this
phase is to deliver the message vector of length $g(K)L$ bits
from the source-clusters towards the destination-clusters in a
multi-hop fashion. By a destination-cluster, we mean a cluster
of size $g(K)$ that includes the corresponding final
destination for a specific source node. We consider a
cluster-based routing scheme in which hop-by-hop communication
between neighboring clusters is performed either horizontally
or vertically as shown in Fig.~\ref{routingg}. Each
intermediate cluster along the path is called a
\emph{relay-cluster}, which acts as a distributed multi-antenna
system to forward its received message vector.

In the first hop, the nodes of the transmitting relay-cluster
\emph{independently} encode the message vector to $\mathcal{C}$
symbols using a Gaussian code-book of power $P^{(2)}=2^\alpha P
g(K)^{\frac{\alpha}{2}-1}$. In Section~\ref{anainter}, we will show
that within each hop $g(K)\mathcal{C}$ symbols are transmitted in
one shot and we benefit from $g(K)$ spatial degrees of freedom in
MIMO communication. These symbols are sent synchronously to the
receiving relay-cluster. The nodes in the receiving relay-cluster
then amplify their receive observation to meet the power requirement
$P^{(2)}$, and forward it to the next cluster along the horizontal
path. Such message propagation continues until the message vector
reaches the cluster with the same vertical boundaries as the
destination cluster, as illustrated in Fig.~\ref{routingg}. Then,
the hop-by-hop distributed MIMO communication among the clusters is
performed vertically until the message vector reaches the
destination cluster. We denote the required time to complete the
routing of a node's message by $T_2$.

%Since there are $g(K)$ nodes in each cluster, the total
%required time for the inter-cluster communication phase equals
%$g(K)T_2$.  Having a large $K$th-level cluster results in
%having a higher achievable inter-cluster transmission rate, and
%fewer number of hops are required for the message to be
%propagated towards the destination. However, having a large
%$K$th-level cluster increases the required number of iterations
%to finish the intra-cluster communication.

\textbf{Phase 3: Message Decoding at the Destination}. In this
phase, each node in the destination cluster has an observation
of the message transmitted by the source $S$. The nodes each
quantize their observation into $Q$ bits and synchronously
transmit the quantized observations towards the destination
using a common constant power $P^{(3)}=P$. This communication
setting can be considered as a simple distributed
Multiple-Input Single-Output (MISO) system in the destination
cluster. We again use a 9-TDMA scheduling in this phase to
determine the active clusters for transmission, which results
in interference mitigation.  We denote the required time to
complete message transmission from the destination cluster to
the destination node by $T_3$.

%Note that by increasing the cluster size $g(K)$, the achievable
%sum-rate only increases in a logarithmical trend, while the
%number of required bits to be delivered to the destination
%increases linearly. Therefore, having a larger $K$th-level
%cluster increases the communication time $T_3$, required in
%this phase. Since there are $g(K)$ nodes in the cluster, the
%total required time for this step equals $g(K)T_3$.

\section{Scaling Laws for Regular Networks}
\label{cresult}

In this section, we derive the asymptotic achievable network
throughput $T(N)$ using the proposed communication scheme.  It
is clear that the efficiency of this scheme hinges on the
optimization of the cluster ($K$th-level cluster) size. While
larger clusters result in higher MIMO gain, better interferer
separation, and lower cluster-hop count, the nodes inside large
clusters need to distribute their messages to a large number of
nodes and compete with many nodes for medium access. By
considering such tradeoffs, we derive the optimal cluster size
to minimize the total transmission time of the three
communication phases. Note that the communication overheads,
such as node scheduling and hierarchy formation, are negligible
in the long run and are ignored in our analysis
\cite{ozgur2}\cite{toumpis}\cite{massimo}.

\subsection{Intra-Cluster Communication} \label{intracomm}

During the $(k+1)$th iteration of intra-cluster message
dissemination, a $k$th-level cluster transmits a message vector to
each of its 8 neighboring $k$th-level clusters in 8 consecutive
$(k+1)$th-level casts. We now derive a lower-bound on the achievable
rate, denoted ${\bf R^k_{\text{Intra}}}$, during each such
transmission.

Consider two adjacent squares of side length $3^{k}$ which
share an edge. The achievable communication rate between the
corresponding two $k$th-level clusters can be expressed using
MIMO capacity results \cite{telatar}. Using a notation similar
to the one we adopted in \eqref{modelformul}, the vector of the
received messages at the $k$th-level receive-cluster can be
expressed as
\begin{equation}
\begin{split}
{\bf Y}_{g(k)\times 1}=H{\bf X}_{g(k)\times 1}+{\bf Z}_{g(k)\times
1}+ {\bf I}_{g(k)\times 1},
\end{split}
\end{equation}
where $H$ represents the $g(k)\times g(k)$ channel matrix.
Treating the interference simply as another source of noise,
the interference plus noise power at a node $j$ in the
receive-cluster is given as $N_0+{\bf I}^k_j$, where ${\bf
I}^k_j$ represents the interference power at $j$ during the
$(k+1)$th-level cast. In order to derive a capacity
lower-bound, we can replace the interference power at all of
the nodes inside a \emph{specific} $k$th-level cluster by
$\mathbb{ I}^k=\max_{j}{\bf I}^k_j$. Assuming that the
transmitted signals, $X_i$, are iid and chosen from an Gaussian
distribution $\mathcal{N}(0,P)$ and perfect channel state
information is available at the receive cluster, the mutual
information between the transmit- and receive-clusters is
lower-bounded by the special case of Gaussian noise plus
interference.  Hence, we have
\begin{equation}
{\bf R^k_{\text{Intra}}}=I({\bf X};{\bf Y}, H)\ge
\log\det(I+\frac{P}{N_0+\mathbb{I}^k} HH^*) ~.
\end{equation}

We now derive a more useful expression to replace the above
general form. An example of intra-cluster MIMO communication is
depicted in Fig.~\ref{sched}(a) for 1st-level clusters of size
$g(1)=9$. The worst case topology happens when the transmitters
are located at the furthest possible distance from the
receivers. Figure~\ref{sched}(b) demonstrates such a topology
in the 3rd-level cast for two horizontally adjacent 2nd-level
clusters of size $g(2)=81$. Without loss of generality, we
label the border (e.g., center red line in
Figure~\ref{sched}(b)) between the two adjacent $k$th-level
clusters as the reference $x$-coordinate.  Then, given the
constraint that there is one node inside each square, setting
the transmit nodes' $x$-coordinates to $x_t=-1, -2, \ldots,
-\sqrt{g(k)}$, and the receive nodes' $x$-coordinates to
$x_r=1, 2, \ldots, \sqrt{g(k)}$ results in maximum separation
between them.

We next find the $y$-coordinates for the worst case topology.
We focus on the transmit and receive nodes which lie in squares
adjacent to the border. The $y$-coordinate for other nodes
follows the same structure by symmetry.
%As shown in
%Figure~\ref{sched}(b), it is clear that the transmit-cluster nodes
%have to lie at $x_t=-1$, and the receive-cluster nodes have to be
%at $x_r=1$.
We index the $y$-coordinates of the horizontal edges from 1 to
$\sqrt{g(k)}+1$ as shown in Fig.~\ref{sched}(b). In general it is
not easy to find the indices along the transmit and receive edges
that minimize ${\bf R^k_{\text{Intra}}}$. However, for our analysis,
since there is a horizontal distance of at least two between any two
nodes chosen from the transmit and the receive clusters (i.e.,
$x_r-x_t\ge 2$), we have $h_{ij}\le \frac{1}{2^{\frac{\alpha}{2}}}$.
For $\alpha>2$, the small value of $h_{ij}$ facilitates tight
lower-bounds for the MIMO communication rate between the two
adjacent clusters. Lemma \ref{lemma1} below states that the
construction of this lower-bound is equivalent to finding the
$y$-coordinate for transmit and receive nodes such that
$\text{Tr}(HH^{*})=\sum_{i\in \mathbb{T}_k}\sum_{j \in
\mathbb{R}_k}|h_{ij}|^2$ is minimized, where $\mathbb{T}_k$
represents the set of nodes in the $k$th-level transmit-cluster and
$\mathbb{R}_k$ is the set of $k$th-level receive-cluster nodes.
Lemma \ref{lemma2} then provides the explicit $y$-coordinates.  The
proofs are given in the appendix.

\newtheorem{lemma}{Lemma}
\begin{lemma}
\label{lemma1} The capacity expression
$\log\det(I+\frac{P}{N_0+\mathbb{I}} HH^*)$ is tightly
lower-bounded by
$\log(1+\frac{P}{N_0+\mathbb{I}}\text{Tr}(HH^{*}))$, where
$\text{Tr}$ is the matrix trace operation.
\end{lemma}

%\begin{IEEEproof}
%Refer to Appendix A.
%\end{IEEEproof}

\begin{lemma}
\label{lemma2}
 The lowest rate in a $(k+1)$th-level cast is achieved when the nodes on the transmit side are located at
 $y_i \in \{1,\cdots,\sqrt{g(k)}+1\} \backslash
\{\frac{\sqrt{g(k)}+1}{2}\}$ and the nodes on the receive side are
located at $y_j \in \{1,\cdots,\sqrt{g(k)}+1\}\backslash
\{\frac{\sqrt{g(k)}+3}{2}\}$ or vice versa due to symmetry.
\end{lemma}

We next use Lemmas \ref{lemma1} and \ref{lemma2} to derive the
achievable communication rate. We define the \emph{vertical
distance} between two nodes $i$ and $j$ as $D_{ij}=|y_i-y_j|$.
Clearly, $D_{ij}\in\{0,1,\cdots,\sqrt{g(k)} \}$. In order to
evaluate $\sum_{i\in \mathbb{T}_k}\sum_{j \in
\mathbb{R}_k}|h_{ij}|^2$ we have to quantify the number of
transmit-receive pairs $(i,j)$, along $x_t=-1$ and $x_r=1$,
which are located at a specific vertical distance $D_{ij}=d$
from each other. We denote this quantity by
$\Phi_d=|\{(i,j)|D_{ij}=d,x_i=-1,x_j=1\}|$.

For a given vertical separation, $D_{ij}=d$, between a pair of
nodes, we have
\begin{equation}
\begin{split}
\sum\sum_{(i,j \text{s.t.}
D_{ij}=d)}|h_{ij}|^2=\sum_{x_t=-\sqrt{g(k)}}^{-1}\sum_{x_r=1}^{\sqrt{g(k)}}\frac{\Phi_d}{\big(\sqrt{(x_r-x_t)^2+d^2}\big
)^\alpha}.\nonumber
\end{split}
\end{equation}
\normalsize

The above sum can be written as $\mathbf{u}(d) \Phi_d$, where
\begin{equation}
\begin{split}
\label{hd}
&\mathbf{u}(d)=\sum_{x_t=1}^{\sqrt{g(k)}}\sum_{x_r=1}^{\sqrt{g(k)}}\frac{1}{\big
(\sqrt{(x_r+x_t)^2+d^2}\big
)^\alpha}>\sum_{x=2}^{\sqrt{g(k)}}\frac{x-1}{(x^2+d^2)^{\frac{\alpha}{2}}}\\&>
\int_{2}^{\sqrt{g(k)}}\frac{x-1}{(x^2+d^2)^{\frac{\alpha}{2}}}dx>\int_{2}^{\sqrt{g(k)}}\frac{x/2}{(x^2+d^2)^{\frac{\alpha}{2}}}dx\\
&=\frac{1}{4(\frac{\alpha}{2}-1)}(\frac{1}{(4+d^2)^{\frac{\alpha}{2}-1}}-\frac{1}{(g(k)+d^2)^{\frac{\alpha}{2}-1}})\quad,
~\alpha>2.
\end{split}
\end{equation}
\normalsize In order to evaluate $\Phi_d$, we first quantify
the number of nodes $\phi_d(q)$, with distance $0\le d \le
\sqrt{g(k)}$ from a node $i$ located at $(x_i=-1,y_i=q)$, $q
\in \{1,\cdots,\sqrt{g(k)}+1\} \backslash
\{\frac{\sqrt{g(k)}+1}{2}\}$, in the $k$th-level
transmit-cluster. It is clear that $\Phi_d=\sum_{q=1,q \ne
\frac{\sqrt{g(k)}+1}{2}}^{\sqrt{g(k)}+1}\phi_d(q)$.

%We have \small
%\begin{equation}
%\nonumber \Phi_d=\sum_{q=1,q \ne
%\frac{\sqrt{g(k)}+1}{2}}^{\sqrt{g(k)}+1}\phi_d(q).\end{equation}
%\normalsize
The following facts are used to compute $\phi_d(q)$:
\begin{itemize}
\item[$\bullet$] Each node in the transmit-cluster (except
    $y_i=\frac{\sqrt{g(k)}+3}{2}$) has 1 node of vertical
    distance 0 from it at the receive-cluster.
\item[$\bullet$] If $q-d\ge 1$ but $q+d>\sqrt{g(k)}+1$, or $q-d<1$ but $q+d\le
\sqrt{g(k)}+1$, there is one node located at distance $d$ from node
$i$.
\item[$\bullet$]If
$q-d \ge 1$ and $q+d \le \sqrt{g(k)}+1$, there are two nodes at
distance $d$ from node $i$.
\item[$\bullet$] For all $q$ and $d$, if
    $y_j=\frac{\sqrt{g(k)}+3}{2}$, then it cannot be used
    in computing $\phi_d(q)$, since by Lemma 2 $y_j$ is removed from the set of
    receiver indices.
\end{itemize}
Hence, we can quantify $\phi_d(q)$ as follows:%

\begin{equation}
\begin{split}
\nonumber &\phi_d(q)=\mathcal{I}\big(q-d\ge 1 \bigcap
q+d>\sqrt{g(k)}+1\big)+\\&\mathcal{I}\big(q-d< 1 \bigcap q+d\le
\sqrt{g(k)}+1\big)+2\mathcal{I}\big(q-d\ge 1 \bigcap q+d\le\\&
\sqrt{g(k)}+1\big)-\mathcal{I}\big(q+d=\frac{\sqrt{g(k)}+3}{2}\bigcup
|q-d|=\frac{\sqrt{g(k)}+3}{2}\big)-\\&\mathcal{I}\big(d=0\big),
\end{split}
\end{equation}
\normalsize where $\mathcal{I}$ represents the indicator function,
such that it equals to 1 if its condition holds and 0 otherwise.

Summarizing the above, we have
\begin{equation}
\label{kdeq} \Phi_d=\left\{\begin{array}{ccc}
\sqrt{g(k)}-1, \quad \text{if} \quad d=0\\
2\sqrt{g(k)}-3, \quad \text{if} \quad d=1\\
2\sqrt{g(k)}-2d-2, \quad \text{if} \quad 2\le d \le \frac{\sqrt{g(k)}-1}{2}\\
2\sqrt{g(k)}-2d, \quad \text{if} \quad d=\frac{\sqrt{g(k)}+1}{2}\\
2\sqrt{g(k)}-2d+2, \quad \text{if} \quad \frac{\sqrt{g(k)}+3}{2}\le
d \le \sqrt{g(k)}
\end{array}\right. .
\end{equation}
Given $\Phi_d$, Lemma \ref{lemma3} below states the scaling law of
$\text{Tr}(HH^{*})$. The proof is given in the appendix.

\begin{lemma}
\label{lemma3}
As $g(k) \rightarrow \infty$, \[\displaystyle
\text{Tr}(HH^{*})=\sum_{i\in \mathbb{T}_k}\sum_{j \in
\mathbb{R}_k}|h_{ij}|^2=\omega(\sqrt{g(k)}).\]
\end{lemma}

Using the above lemmas, we characterize the intra-cluster
communication time. During the $k$th iteration, we denote by
$l_a[k]$ the minimum spatial separation between the interferers
which are located at tier $a$ and a node inside the receiving
cluster. Because of the 9-TDMA transmission scheme, there are
at most $\lceil\frac{1}{3}\sqrt{\frac{N}{g(k)}}\rceil$ tiers
during the $k$th iteration. In order to obtain an upper-bound
on the interference, we assume that all of the $g(k)$ nodes in
an interferer cluster are located at the closest boundary to
the receive-cluster R. Therefore, $l_a[k] \ge
(3a-2)\sqrt{g(k)}$. Hence, the interference is upper-bounded as

\begin{equation}
\begin{split}
\label{intform} \mathbb{I}^k&\le
\sum_{a=1}^{\lceil\frac{1}{3}\sqrt{\frac{N}{g(k)}}\rceil}8a\frac{Pg(k)}{l_a[k]^\alpha}\le
\frac{8P}{g(k)^{\frac{\alpha}{2}-1}}\sum_{a=1}^{\lceil\frac{1}{3}\sqrt{\frac{N}{g(k)}}\rceil}\frac{a}{(3a-2)^\alpha}\\
&<
\frac{8P}{g(k)^{\frac{\alpha}{2}-1}}\big(1+\frac{1}{3}\sum_{a=1}^{\lceil\frac{1}{3}\sqrt{\frac{N}{g(k)}}\rceil-1}
\frac{1}{(3a+1)^{\alpha-1}}
\big)\le\frac{8P}{g(k)^{\frac{\alpha}{2}-1}}\\&\big(1+\frac{1}{3}\int_{a=0}^{\lceil\frac{1}{3}\sqrt{\frac{N}{g(k)}}\rceil-1}
\frac{1}{(3a+1)^{\alpha-1}}da \big)\xrightarrow[N\rightarrow
\infty]{}\frac{8c_2P}{g(k)^{\frac{\alpha}{2}-1}},
\end{split}
\end{equation}
\normalsize where $c_2$ is a constant.

Eqn.~\eqref{intform} suggests the following.  By increasing the
size of the clusters, due to the scheduling algorithm, the
spatial separation of the interferer nodes from the
receive-cluster increases. Although the number of nodes inside
each cluster is increasing, the overall effect of interference
diminishes. Hence, in the limit, for large clusters, the
communication becomes noise limited. Therefore, we have the
following conclusion.
\newtheorem{proposition}{Proposition}
\begin{proposition} \label{prop1}
As $g(k) \rightarrow \infty$, during a $(k+1)$th-level cast,
\[
{\bf R^k_{Intra}} = \omega(\log{g(k)}) ~.
\]
\end{proposition}
\begin{IEEEproof}
The proof immediately follows from Lemmas \ref{lemma1} and
\ref{lemma3} and the fact that the interference power tends to
0 as $g(k) \rightarrow \infty$.
\end{IEEEproof}

Since the proposed intra-cluster communication uses $K$
iterative steps to distribute all $g(K)$ bits of the source $S$
to the nodes inside the $K$th-level cluster, the time required
to finish this phase is
\begin{equation}
\begin{split}
\vspace{-5mm}
 T_1=\sum_{k=1}^{K}T^k_1,
\vspace{-5mm}
\end{split}
\end{equation}
where $T^k_1$ represents the time taken to finish the $k$th
iteration. In the $k$th iteration, 9 consecutive $k$th-level
casts are performed, each with a rate of ${\bf
R^{(k-1)}_{Intra}}=\omega(\log{g(k-1)})$. At the same time that
MIMO transmission is in effect between two neighbor
$(k-1)$th-level clusters of size $g(k-1)$, the $(k-1)$th-level
transmit-cluster receives new packets to transmit them to
another adjacent cluster of size $g(k-1)$ during the next MIMO
transmission. The required time to finish the $k$th iteration
is, therefore, equal to the maximum of these two times. Since
$g(k-1)L$ bits are sent in one shot during each such
transmission using MIMO communication, we have
\begin{equation}
\label{time} T^k_1\le\max(9\frac{g(k-1)L}{\log{g(k-1)}},9T^{k-1}_1)<
9c_3Lg(k-1),
\end{equation}
for a constant $c_3$ and $k>1$. This can be easily shown by
induction since $T^1_1$ is a constant.  This leads to the
following conclusion.

\begin{proposition}
The total required time to finish the intra-cluster
communication phase is $o(g^2(K))$.
\end{proposition}
\begin{IEEEproof}
Since $g(k)=3^{2k}$, the time required to finish the
broadcasting of each node's packet is upper bounded as
\begin{equation}
\begin{split}
\label{time2} T_1<
9c_3L\sum_{k=1}^{K}3^{2(k-1)}=9c_3L\frac{1}{8}(g(K)-1) ~.
\end{split}
\end{equation}
Hence, $T_1=o(g(K))$.  Since this step has to be repeated
$g(K)$ times for all of the nodes inside each $K$th-level
cluster, the total required time is $o(g^2(K))$.
\end{IEEEproof}

\subsection{Inter-Cluster Communication}
\label{anainter} 
\begin{figure}
  \centering
\subfigure{\includegraphics[scale=.55]{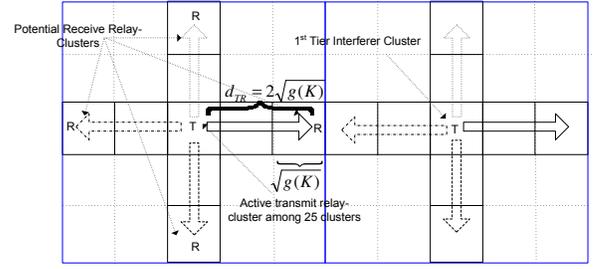}}
  \caption{Alternate inter-cluster communication scheme and interferer locations.}\label{intra}
\vspace{-3mm}
 \end{figure}

We derive a lower bound for the achievable rate,
denoted ${\bf R_{\text{Inter}}}$, for the second phase of the
proposed scheme, by considering an alternate inter-cluster
communication scheme as shown in Fig.~\ref{intra}, where we assume a
one-cluster separation between each pair of transmit-receive relay
clusters. The capacity result obtained by this alternate scheme can
serve as a lower bound for the original scheme because the distance
between any two nodes within the alternate scheme is larger than
that in the old scheme.

Furthermore, we observe the following about the alternate
scheme:
\begin{itemize}

\item[$\bullet$] The $a$th-tier interferer nodes are
    located at a distance of at least $(4a-2)\sqrt{g(K)}$
    from the destination cluster, for $1 \le a \le
    \lceil\frac{\sqrt{N}}{5\sqrt{g(K)}}\rceil$.

\item[$\bullet$] A 25-TDMA scheme is used, so there is one active
transmit relay-cluster among 25 clusters of size $g(K)$ (the
optimal cluster size).
\end{itemize}
In the following we quantify the interference in each cluster,
and demonstrate that $g(K)$ degrees of freedom are achievable
in the proposed MIMO communication.

The derivation of the maximum amount of interference for the nodes
of each receive cluster follows the same lines as~(\ref{intform}),
and we only replace the spatial separation of $3a-2$ with $4a-2$. It
can be easily verified that as before, by replacing this new value,
the term inside the summation in~(\ref{intform}) approaches a
constant, and we can upper-bound the inter-cluster interference as
$\mathbb{I}_{\text{Inter}}\le\frac{8c_4P^{(2)}}{g(K)^{\frac{\alpha}{2}-1}}=8
c_4 2^{\alpha}P$ for a constant $c_4$.  This leads to the following
two lemmas, which were developed in \cite{ozgur2} for a dense
network but here we show to hold in the extended network under
consideration.  The proofs are given in the appendix.

\begin{lemma}
\label{lemma4} The achieved mutual information between any pair of
transmit-receive relay-clusters during the multi-hop phase, and
hence ${\bf R_{\text{Inter}}}$, grows at least linearly with $g(K)$.
\end{lemma}

\begin{lemma}
\label{lemma5} There exists a strategy to encode the observation at
each node in the final receive-cluster at a fixed rate, $Q$, and
still maintain the above linear growth of the mutual information for
the $g(K)\times g(K)$ quantized MIMO channel.
\end{lemma}

We now summarize the overall required time to complete this
phase in the following proposition:
\begin{proposition}
\label{reg-interc} The total required time to complete the
inter-cluster communication phase is $o(\sqrt{Ng(K)})$.
\end{proposition}
\begin{IEEEproof}
During each step of the proposed 25-TDMA communication,
$\frac{1}{25}$ of the clusters act as active sources. Due to
Lemmas \ref{lemma4} and \ref{lemma5}, the full spatial
multiplexing is achieved in this phase, and at any point along
the route the rate between immediate adjacent clusters is
lower-bounded by the rate between the clusters located two
cluster-hops away. Since $\mathcal{C}$ symbols are sent over
each cluster-hop, the communication time between each pair of
transmit-receive relay-clusters scales as
$T_2=\frac{\mathcal{C}g(K)}{g(K)}=\mathcal{C}$.

There are $O(\sqrt{\frac{N}{g(K)}})$ cluster-hops between the
source and the destination since there are at most a total of
$\sqrt{\frac{N}{g(K)}}$ squares of size $g(K)$ along a
horizontal path as well as a vertical path. Therefore, the
end-to-end communication time is upper-bounded by $25
\mathcal{C}\sqrt{\frac{N}{g(K)}}$. We also should note that
there are $g(K)$ nodes inside each cluster.  Therefore, the
total time required to complete the inter-cluster communication
is upper-bounded by
$25g(K)\mathcal{C}\sqrt{\frac{N}{g(K)}}=25\mathcal{C}\sqrt{Ng(K)}$.
\end{IEEEproof}

\subsection{Message Decoding at the Destination}
\label{mesdeco}

%In this phase the message intended for each destination has
%arrived at its corresponding cluster and is distributed among
%the $g(K)$ nodes of the destination-cluster. These nodes act as
%a virtual MISO system and send their corresponding observation
%to the destination. Based on the result of Lemma \ref{lemma5},
%each node quantizes its $\mathcal{C}$ observations to
%$\mathcal{C}Q$ bits and send it towards the destination.

Let $\mathbb{I}_{\text{D}}$ represent the interference power at
any destination node during this phase.  Following our analysis
of interference in~(\ref{intform}), and due to the 9-TDMA used
to select the active $K$th-level clusters, it is clear that
$\mathop{\lim}\limits_{g(K)\rightarrow
\infty}\mathbb{I}_{\text{D}}=0$.  Hence, the achievable
communication rate can be written as
\[
{\bf
R_{\text{D}}}=I({\bf X};{\bf y},
H)=\log(1+\frac{P}{N_0}\sum_{l=1}^{g(K)}|h_{lD}|^2) ~,
\]
where
$|h_{lD}|^2$ is the square amplitude of the channel between any
node $l$ in the cluster and the destination $D$, and ${\bf y}$
is the vector of the received message at the destination.

In order to derive a lower-bound on ${\bf R_{\text{D}}}$, we
assume that the destination is located on one of the corners of
the cluster. In a regular network, this choice results in the
smallest value for $\sum_{l=1}^{g(K)}|h_{lD}|^2$. To evaluate
this term, we use Fig.~\ref{sched}(a) as an example. We
consider the node located at the upper-right corner. There are
3 nodes in 3 squares which are adjacent to the square
containing this node, each located at a distance of at most
$2\sqrt{2}$ from the destination. We call these nodes 1st-tier
nodes. Likewise, the 2nd-tier consists of 5 nodes with a
distance of at most $3\sqrt{2}$ from the destination.
Generalizing the idea, it can be verified that in the $z$th
tier, there are $2z+1$ nodes of distance at least
$\sqrt{2}(z+1)$ from the destination $D$. Therefore, we have

\begin{equation}
\begin{split}
&\lim_{g(K) \rightarrow \infty}\sum_{l=1}^{g(K)}|h_{lD}|^2>
\sum_{z=1}^{\infty}\frac{2z+1}{(\sqrt{2}(z+1))^\alpha}=\\&\sum_{z=2}^{\infty}\frac{1}{\sqrt{2}^\alpha}(\frac{2}{z^{\alpha-1}}-\frac{1}{z^{\alpha}})=
\frac{1}{\sqrt{2}^\alpha}(2\xi(\alpha-1)-\xi(\alpha)-1)=c_6,\nonumber
\end{split}
\end{equation}
\normalsize where $\xi$ is the Riemann-Zeta function. Hence,
${\bf R_{\text{D}}}$ is a constant $c_6$ and does not scale
with the cluster size.  This leads to the following conclusion.

\begin{proposition}
The required time to complete the message decoding phase is
$o(g^2(K))$.
\end{proposition}
\begin{IEEEproof}
Based on the result of Lemma 5 each node quantizes its observation
of the message using a constant, $Q$, bits and the communication
rate is lower-bounded by a constant, therefore, the required time to
finish each MISO transmission to deliver $g(K)$ bits is $T_3<
\frac{\mathcal{C}Qg(K)}{c_6}$. Because of the 9-TDMA scheme and the
fact that there are $g(K)$ destination nodes inside each cluster,
the overall required time to complete this step is $o(g^2(K))$.
\end{IEEEproof}

\subsection{Constructive Lower-Bound on the Capacity}
\label{capreg} The lower bound on the capacity can now be derived,
since we know the scaling behavior of the required time for each
phase and the total number of bits transmitted in the network. As
$g(K)\rightarrow \infty$, the total required time to finish
transmitting a packet using the proposed communication scheme is
\begin{equation}
\begin{split}
\nonumber
\text{t}_{\text{total}}&=o(g(K)^2)+o(\sqrt{Ng(K)})+o(g(K)^2)\\&=o(\sqrt{Ng(K)})+o(g(K)^2).
\end{split}
\end{equation}

Therefore, since each node sends a packet of length $g(K)$ and there
are $N$ nodes in the network, the overall network capacity can be
written as
\begin{equation}
\begin{split}
T(N)=\frac{Ng(K)}{\text{t}_{\text{total}}}=
\frac{Ng(K)}{o(\sqrt{Ng(K)})+o(g(K)^2)}.\nonumber
\end{split}
\end{equation}
 Note that
since we are interested in the scaling behavior of the capacity, our
goal is to find $K^*=V(N)=\arg\max_{K}T(N)$ for some function $V$.
By letting $o(\sqrt{Ng(K)})=o(g(K)^2)$, we find the optimal
$g(K^{*})=c_7 N^{\frac{1}{3}}$, for some constant $c_7$.
Substituting this into $T(N)$, we have the following main result.
\newtheorem{theorem}{Theorem}
\begin{theorem}
Assuming $\alpha>2$, the achievable network throughput for all
source-destination pairs using the proposed algorithm is
$T(N)=\omega(N^{\frac{2}{3}})$ with a power scaling of
$P^{(2)}=2^\alpha P g(K)^{\frac{\alpha}{2}-1}=c_8
N^{\frac{\alpha}{6}-\frac{1}{3}}$ for some constant $c_8$.
\end{theorem}

\newtheorem{corollary}{Corollary}

\section{Scaling Laws for Random Networks}
\label{numsim}

\subsection{Random Network with Unit Node Density} \label{unit}
In this section, we assume that $N$ nodes are distributed
uniformly randomly in the area $B_N$ with unit density and study the achievable
network throughput. The communication algorithm is in nature the
same as that of the case for a regular network.  We explain the
minor differences where appropriate and derive the asymptotic
throughput for this random network.

\subsubsection{Communication inside Squares} In a random network,
we cannot guarantee that each unit area contains exactly one node.
Hence, instead of squares with unit area, we study squares of area
$A\log{N}$, for $A>1$, and find a tight bound on the number of
nodes inside each such square as stated in Lemma \ref{lemma6}. We
then apply the results of Section~\ref{cresult} to this setting.

\begin{lemma} \label{lemma6}
For all squares of area $A\log{N}$, $A>1$, indexed as $1 \le j
\le \lfloor\frac{N}{A\log{N}}\rfloor$, the number of nodes,
$M_j$, inside each square is in the interval
$[(1-\delta)A\log{N},(1+\delta)A\log{N}]$, $\forall \delta>0$,
with probability $p \rightarrow 1$.
\end{lemma}

The proof is given in the appendix.

We assume that instead of packets of length $g(K)L$, the nodes
transmit their messages in packets of length $ g'(K)L A\log{N}$ bits
for the case of random networks.  Here $g'(K)$ represents the
optimal number of squares with area $A\log{N}$ inside the cluster.
For this topology, we add an extra phase before each 1st-level cast
of the intra-cluster communication in order to distribute each
node's message inside the squares of area $A\log{N}$.

In this phase, each source node transmits a different sub-block of
$L$ bits to each of the $A\log{N}-1$ other nodes inside the square.
As before, a 9-TDMA scheme is used to determine the set of active
squares. Within each active square of area $A\log{N}$, only one node
sends its message at a given time. Therefore, using a similar
analysis to~(\ref{intform}), it is straightforward to deduce that
the interference tends to 0 inside each square and the communication
is limited by noise as $N\rightarrow \infty$.

The distance between any pair of $A\log{N}$ nodes located
inside each square is at most $\sqrt{2A\log{N}}$. Therefore,
the communication rate between each pair is lower-bounded by
$\log{(1+P(2A\log{N})^{- \frac{\alpha}{2} } })\rightarrow
P(2A\log{N})^{-\frac {\alpha}{2} }$ as $N\rightarrow \infty$.
Hence, each node can distribute its packet of length $AL\log{N}
$ bits among its neighbor nodes in the square in $c_9
(\log{N})^{1+\frac{\alpha}{2}}$ seconds for some constant
$c_9$.

\subsubsection{Modified Three-phase Scheme and Capacity Evaluation}
The same scheme for intra-cluster communication can now be used
with a slight modification, such that instead of having one node
in each 0th-level cluster, we now have $A\log{N}$ nodes. In order
to find the topology leading to the lowest communication rate, we
follow the lines of analysis in Section~\ref{cresult}-A with the
slight difference that during the $k$th iteration, we assume all
$A\log{N}$ nodes inside each square of area $A\log{N}$ are
co-located at the worst-case point, whose location can be found
similarly to how Lemma 2 is derived. In order to derive the
lower-bound on the communication rate in this case, we also need
to consider the fact that the distance between any two nodes is
multiplied by $\sqrt{A\log{N}}$ compared to the case in a regular
network. It is clear that since the square area grows with $N$,
the resulting interference, $\mathbb{I}^k_{\text{random}}$, during
the $k$th iteration obeys $\mathbb{I}^k_{\text{random}}\rightarrow
0$.

Applying the above changes to the number of nodes in each
square and their relative distance, the result of Proposition
\ref{prop1} is modified to the following.
\begin{proposition}
In the random network with unit node density, as $g'(k)\rightarrow
\infty$,
\[
{\bf R^{k}_{Intra}} = \omega( \log{( 1+\frac{g'(k)(A\log{N})}{
(A\log{N})^{ \frac{\alpha}{2} } } ) } ) ~.
\]
\end{proposition}
Hence, if $g'(k)=o((\log{N})^{\frac{\alpha}{2}-1})$, then
\[
{\bf R^{k}_{Intra}}=\omega(g'(k)(\log{N})^{1-\frac{\alpha}{2}}) ~;
\]
otherwise,
\[
{\bf R^{k}_{Intra}}=\omega(\log{ (g'(k)(\log{N})^{
1-\frac{\alpha}{2} })}) ~.
\]

Defining ${T^{'}}^k_1$ similarly to ${T}^k_1$ in
Section~\ref{cresult}-A, we have the required time to finish
the $k$th iteration
\begin{equation}
\begin{split}
&T^{k}_{\text{tot}}\le c_9
(\log{N})^{1+\frac{\alpha}{2}}+{T^{'}}^k_1=c_9
(\log{N})^{1+\frac{\alpha}{2}}\\&+\max(9\frac{AL\log{N}
g'(k-1)}{{\bf R^{(k-1)}_{Intra}}},9{T^{'}}^{k-1}_1)\le
c_{10}(\log{N})^{\frac{\alpha}{2}} g'(k).\nonumber
\end{split}
\end{equation}
The first term in the above summation accounts for the required time
to propagate the message inside the squares of area
$\sqrt{A\log{N}}$. The second term has the same structure
as~\eqref{time}. The last inequality follows from induction and the
fact that ${T^{'}}^1_1\le c_{10}(\log{N})^{\frac{\alpha}{2}}$. From
this, we obtain the following conclusion.
\begin{proposition}
The total time required to complete the intra-cluster
communication phase in a random network with unit node density is
$o((\log{N})^{\frac{\alpha}{2}+1}{g'}^2(K))$.
\end{proposition}
\begin{IEEEproof}
Along the same line as~(\ref{time2}), and by replacing
$T_{\text{tot}}^k < c_{10}(\log{N})^{\frac{\alpha}{2}}
g'(k)$, it is clear that%
\[
T^{'}_1=\sum_{k=1}^{K}T_{\text{tot}}^k=o((\log{N})^{\frac{\alpha}{2}}
g'(K)) ~.
\]
Since there are $g'(K) A\log{N} $ nodes inside each cluster,
the overall required time to finish this phase is
$o((\log{N})^{\frac{\alpha}{2}+1} {g'}^2(K))$.
\end{IEEEproof}

The inter-cluster communication and message-decoding phases are
almost identical to the regular network case. At the beginning
of the inter-cluster communication, within each cluster of size
$g'(K)A\log{N}$ there are $A g'(K)\log{N}$ nodes. By using
Lemmas \ref{lemma4} and \ref{lemma5}, it is clear that the
achieved capacity in this phase, ${\bf R_{Inter}}$, grows at
least linearly with $g'(K)\log{N} $ and the power constraint in
this phase is $P^{(2)}=2^\alpha P
(g'(K)\log{N})^{\frac{\alpha}{2}-1}$.  Hence, we have
\begin{proposition}
In the random network with unit node density the total required
time to complete inter-cluster communication is
$o(\sqrt{Ng'(K)\log{N}})$.
\end{proposition}

The required time for decoding at the destination cluster
follows from the analysis in Section~\ref{mesdeco}. It can be
shown that we can maintain a constant rate ${\bf R_D}$ of
communication, and therefore the required time to complete MISO
communication is $T^{'}_3=\frac{\mathcal{C}Qg'(K)\log{N}} {{\bf
R_D}}=o(g'(K)\log{N} )$.  Hence, we have
\begin{proposition}
In the random network with unit node density the total required
time to complete the message decoding phase is $o((\log{N})^2
{g'}^2(K) )$.
\end{proposition}

We can now write the total communication time for a random
network as %
\[
\text{t}^{'}_{\text{total}}=o(\sqrt{N\log{N}g'(K)})+o((\log{N})^{\frac{\alpha}{2}+1}{g'}^2(K)) ~,
\]
and the network throughput is given by
\[
T(N)= \frac{
N(g'(K)\log{N})}{o(\sqrt{N\log{N}g'(K)})+o((\log{N})^{\frac{\alpha}{2}+1}
g'(K)^2)} ~.
\]
To maximize $T(N)$, we set
$o(\sqrt{N\log{N}g'(K)})=o((\log{N})^{\frac{\alpha}{2}+1}
{g'}^2(K))$. Therefore, the optimal cluster size $g'(K)\log{N}$ is
$c_{11}N^{\frac{1}{3}}(\log{N})^{\frac{2-\alpha}{3}}$, for a
constant $c_{11}$.  Hence, we obtain the following main result for a
random network.
\begin{theorem}
In the random network with unit node density assuming $\alpha> 2$,
the achievable network throughput for all source-destination pairs
using the proposed algorithm is $T(N)=\omega
\big(N^{\frac{2}{3}}(\log{N})^{\frac{2-\alpha}{6}}\big)$ with a
required power scaling of $P^{(2)}=2^\alpha P
(g'(K)\log{N})^{\frac{\alpha}{2}-1}=c_{12}N^{\frac{\alpha}{6}-
\frac{1}{3}}(\log{N})^{-\frac{(\alpha-2)^2}{6}}$ for some constant
$c_{12}$.
\end{theorem}

\subsection{Random Network with Density $\lambda=\Omega(\log{N})$}

We can use a similar algorithm as Section~\ref{unit} to find
constructive lower-bounds on the capacity scaling behavior of an
extended network with a node density $\lambda=\Omega(\log{N})$.

\begin{lemma}
For all squares of unit area indexed as $1 \le j \le N$, if
$\lambda=\Omega(\log{N})$, the number of nodes, $M_j$, inside the
square is in the interval $[(1-\delta)\lambda,(1+\delta)\lambda]$,
$\forall \delta$, with probability $p \rightarrow 1$.
\end{lemma}

\begin{proof}
The proof is similar to the proof of Lemma~\ref{lemma6}. The
probability of lying in a square of unit area equals
$P(S_i=1)=\frac{1}{N}$. Here, $S=\sum_{i=1}^{\lambda N}S_i$
represents the number of nodes inside a unit square and we have
$p[S<(1-\delta)\lambda N
P(S_i=1)]=p[S<(1-\delta)\lambda]<e^{-\lambda\frac{\delta^2}{2}}$.
Using the upper-bound version of the bound we have,
$p[S>(1+\delta)\lambda N
P(S_i=1)]=p[S>(1+\delta)\lambda]=e^{-\lambda f(\delta)}$. If we
choose $\lambda \ge A\log{N}$ for
$A>\max(\frac{2}{\delta^2},\frac{1}{f(\delta)})$, with probability
1 the number of nodes $M_j$ inside all unit squares follows
$(1-\delta)\lambda<M_j<(1+\delta)\lambda$, $ 1\le j\le N$ and
$\forall \delta>0$.
\end{proof}

As opposed to squares of area $A\log{N}$ in the previous section,
the base squares of unit area are considered and the nodes
transmit their messages in packets of length $\lambda L g(K)$. By
employing the 9-TDMA communication, interference tends to a
constant inside each unit square. The distance between any
pair of $\lambda$ nodes located inside each unit square, is at
most $\sqrt{2}$. Therefore, a constant communication rate between
each pair is achieved inside the square, and each node can
transmit $\lambda L$ bits to its neighbor nodes in $c_{13}\lambda$
seconds for some constant $c_{13}$

 We replicate the analysis done in~\eqref{intform} and consider the fact that instead of $g(k)$ nodes, the $k$th-level cluster
 consists of $\lambda g(k)$ nodes. Therefore, by
modifying~\eqref{intform} we can show that the interference
$\mathbb{I}^k_{\text{random}}$ during the $k$th iteration obeys
\begin{equation} \label{intran2} \mathbb{I}^k_{\text{random}}\le
\frac{8Pc_2\lambda}{g(k)^{\frac{\alpha}{2}-1}}.
\end{equation}
 If $g(k)=o(\lambda^{\frac{2}{\alpha-2}})$
(high interference regime) the rate of the $k$th-level cast scales
as $\omega(\log{(g(k))})$, and if
$g(k)=\omega(\lambda^{\frac{2}{\alpha-2}})$, since
$\lim_{g(k)\rightarrow
\infty}\mathbb{I}^k_{\text{random}}\rightarrow 0$ the rate scales
as $\omega(\log{(\lambda g(k))})$.

In the $k$th iteration, a $(k-1)$-level cluster transmits
$3^{2(k-1)}\lambda L$ bits to one of its neighbors, and using the
same line of reasoning that we used to derive $T^k_{\text{tot}}$
in Section~\ref{unit}, we have
\begin{equation}
\begin{split} T^k_{\text{tot}}&\le c_{13}\lambda+{T^{'}}k \\&\le
c_{13}\lambda+\max(9\frac{\lambda L
g(k-1)}{\log{g(k-1)}},9{T^{'}}^{k-1}_1)\le 9c_{14}\lambda L
g(k-1). \nonumber
\end{split}
\end{equation}

\begin{proposition}
 The total time required to finish the intra-cluster
communication phase in a random network with density
$\lambda=\Omega(\log{N})$ is $o(\lambda^2 g^2(K))$.
\end{proposition}
\begin{proof}
Along the same line as~(\ref{time}), and by replacing
$T_{\text{tot}}^k < 9 c_{14}\lambda L g(k-1)$, it can be deduced
that $T_1=o(\lambda g(K))$. There are $\lambda g(K)$ nodes inside
each cluster of size $g(K)$, therefore, the overall required time
to finish this phase is $o(\lambda^2 g^2(K))$.
\end{proof}

The achieved capacity in the inter-cluster phase, ${\bf R_{Inter}}$,
grows at least linearly with $\lambda g(K)$ and the power constraint
in this phase is $P^{(2)}=2^\alpha P g(K)^{\frac{\alpha}{2}-1}$.

\begin{proposition}
The total required time to finish the inter-cluster communication
(routing) in a random network with node density
$\lambda=\Omega(\log{N})$ is $o(\sqrt{Ng(K)}\lambda)$.
\end{proposition}
\begin{proof}
The proof follows from the proof of Proposition~\ref{reg-interc}
and the fact that there are $\lambda g(K)$ nodes in each cluster
instead of $ g(K)$ nodes.
\end{proof}

To evaluate the required time for message decoding at the
destination cluster, it can be easily shown that in the MISO
analysis, the lower-bound on the sum $\sum_{l=1}^{\lambda
g(K)}|h_{ld}|^2$ equals $c_{15} \lambda$, since in the worst case
scenario, in tier $z$ all the $(2z+1)\lambda$ nodes are located at
the distance $(z+1)\sqrt{2}$ from the destination.

For the interference analysis, considering~\eqref{intran2} and
replacing $k=K$, we again have two paradigms. If
$\lambda=o(g(K)^{\frac{\alpha}{2}-1})$, $\mathbb{I_D}\rightarrow
0$, otherwise $\mathbb{I_D}=o( \frac{\lambda}{
g(K)^{\frac{\alpha}{2}-1} })$. The rate of communication in the
first case obeys

\begin{equation}
\nonumber
{\bf R_D} \le \log(1+P \frac{c_{15}\lambda}{N_0}),
\end{equation}

 and in
the second case it obeys

\begin{equation}
{\bf R_D} \le \log(1+\frac{c_{15}\lambda}{ \frac{\lambda}{
g(K)^{\frac{\alpha}{2}-1} }})=o(\log{g(K)}). \nonumber
\end{equation}

 In both cases,
$T_3=\frac{\lambda g(K)\mathcal{C}_2 }{{\bf R_D}}=o(\lambda
g(K))$. Hence,

\begin{proposition}
The total required time to finish the message decoding phase in a
random network with node density $\lambda=\Omega(\log{N})$ is
$o(\lambda^2 g(K)^2 )$.
\end{proposition}

We can now write the total communication time for this setting as
$\text{t}^{'}_{\text{total}}=o(\sqrt{Ng(K)}\lambda)+o(\lambda^2g(K)^2)$.
The network throughput in this case follows
\begin{equation}
 T(N)= \frac{(\lambda
N)(g(K)\lambda)}{o(\sqrt{Ng(K)}\lambda)+o(\lambda^2 g(K)^2)}.
\nonumber
\end{equation}
 To
maximize the throughput we set $o(\sqrt{Ng(K)}\lambda)=o(\lambda^2
g(K)^2)$. Therefore, the cluster size in this case can be
expressed as
$c_{16}\frac{N^{\frac{1}{3}}}{\lambda^{\frac{2}{3}}}$. The optimal
cluster size decreases by increasing the node density, which is
expected since a cluster of area $g(K)$ has more nodes compared to
the case of regular networks due to higher node density and as a
result the same degrees of freedom can be achieved in a smaller
area.
\begin{theorem}
The achievable network throughput assuming $\alpha> 2$ for all
possible source-destination pairings and using the proposed
algorithm for random networks with $\lambda=\Omega(\log{N})$
follows $T(N)=\omega((\lambda N)^{\frac{2}{3}})$ with a required
power scaling of $P^{(2)}=2^\alpha P
g(K)^{\frac{\alpha}{2}-1}=c_{17}\frac{N^{\frac{\alpha}{6}-\frac{1}{3}}}{\lambda^{\frac{\alpha}{3}-\frac{2}{3}}}$.
\end{theorem}

\section{Conclusion}
 \label{conc}
In this paper we study the communication among nodes in an
extended network, where both signal interference and power
decay are limiting factors of the network capacity. Through the
asymptotic analysis of a proposed three-phase communication
scheme, we quantify how the appropriate combination of node
cooperation, in the form of distributed virtual antenna arrays,
and multi-hop message relaying, among optimally sized clusters
of nodes, can increase the network throughput.  The use of
spatially separated clusters of distributed antennas brings
benefits from both the spatial multiplexing gain of MIMO and
the mitigation of interference. At the same time, multi-hop
communication among the clusters allows significant power
scaling advantage over direct MIMO communication between the
source and destination clusters.  The derived asymptotic
throughput of the proposed communication scheme provides a
constructive lower bound to the capacity of an extended network
with node cooperation.

\begin{appendices}
\section{Proof of Lemma 1}
From linear algebra it is well known that the derivative of the
determinant of a square matrix, $A$, can be written as $d
\det(A)=\text{Tr} (\text{adj}(A)dA)=\det(A)tr (A^{-1}dA)$ for an
invertible matrix $A$. Using Taylor series expansion, we have
$\det(A+\delta X)= \det(A)+\text{Tr}
(\text{adj}(A)X)\delta+O(\delta^2)$, for small $\delta$ and matrix
$X$. The last term is obtained by computing the second order
derivative of the determinant. Therefore, we can simplify the
argument of the $\log$ function in the given capacity expression as
follows, $\det(I+\frac{P}{N_0+\mathbb{I}} HH^*)=\det(I)+
\det(I)\text{Tr} (I^{-1}\frac{P}{N_0+\mathbb{I}} HH^*) +O(\delta^2)=
1+ \text{Tr}( \frac{P}{N_0+\mathbb{I}} HH^*)+O(\delta^2)$. Since
$h_{ij}<\frac{1}{2^{\frac{\alpha}{2}}}$, we can choose
$\delta=\frac{1}{2^\alpha}$.

\section{Proof of Lemma 2}
There are $\sqrt{g(k)}$ transmitter nodes (at $x_t=-1$) and
$\sqrt{g(k)}$ receiver nodes (at $x_r=1$), and $\sqrt{g(k)}+1$
horizontal locations for the nodes to be placed. Therefore, we
need to choose one of the indices on each side and remove it
from the set of node locations. To minimize $\sum_{i\in
\mathbb{T}_k}\sum_{j \in \mathbb{R}_k}|h_{ij}|^2$, we should
choose the index $i$ on the transmit side and $j$ on the
receive side for which the sums $\sum_{g \in
\mathbb{R}_k}|h_{ig}|^2$ and $\sum_{g \in
\mathbb{T}_k}|h_{gj}|^2$ are maximized, and remove them from
the index set.

The index $i$ chosen to be removed from the transmitter set is
the one which results in the largest number of receiver indices
with small vertical distance from node $i$. It is possible for
at most two distinct nodes with indices in the receive-cluster
to have a common distance from a node with index $i$ in the
transmit-cluster. Choosing $i$ such that
$y_i=\frac{\sqrt{g(k)}+1}{2}$ or $y_i=\frac{\sqrt{g(k)}+3}{2}$,
results in having a node which has the largest number of
receiver nodes with small distances from it. (For example, this
is equivalent to choosing $y_i=5$ or $y_i=6$ in
Fig.~\ref{intra1} (b)). Let's set $i=\frac{\sqrt{g(k)}+1}{2}$.
By symmetry, the removed index in the receive-cluster equals to
one of the above given indices. Choosing
$j=\frac{\sqrt{g(k)}+3}{2}$ instead of
$j=\frac{\sqrt{g(k)}+1}{2}$ results in removing channels with
larger amplitudes from $\sum_{g \in \mathbb{T}_k}|h_{gj}|^2$.
Swapping the values of $i$ and $j$ results in the same
achievable rate.

\section{Proof of Lemma 3}
We evaluate $\mathop{\lim}\limits_{g(k)\rightarrow
\infty}\sum_{d=0}^{\sqrt{g(k)}}\mathbf{u}(d)\Phi_d$. For
${g(k)\rightarrow \infty}$, the second term in~\eqref{hd}
approaches 0 and we have $\lim\limits_{g(k)\rightarrow
\infty}\mathbf{u}(d)\rightarrow
\frac{c_1}{(4+d^2)^{\frac{\alpha}{2}}}$. Using~\eqref{kdeq},
for $d>0$, $\Phi_d\ge 2\sqrt{g(k)}-2d-2$. Therefore, we have $
\lim_{g(k)\rightarrow
\infty}\sum_{d=0}^{\sqrt{g(k)}}\mathbf{u}(d)\Phi_d \ge
\lim_{g(k)\rightarrow \infty}\sum_{d=0}^{\sqrt{g(k)}}
\frac{c_1}{ (4+d^2)^{\frac{\alpha}{2} }} \Phi_d>
 \frac{c_1}{2^\alpha}(2\sqrt{g(k)}-2)+2\sqrt{g(k)}\sum_{d=1}^{\sqrt{g(k)}}\frac{c_1}{
(2+d)^{{\alpha} }}-J\rightarrow C\sqrt{g(k)}$, where the last
equality follows from the fact that
$\sum_{d=1}^{\sqrt{g(k)}}\frac{-2d-2}{(d+2)^\alpha}$ in limit
tends to a constant, $J$,  and
$\sum_{d=1}^{\sqrt{g(k)}}\frac{c_1}{ (2+d)^{{\alpha} }}$ can be
written in terms of the Riemann-Zeta function with a constant
argument. Therefore, the final result holds for some constant
$C$.

\section{Proof of Lemma 4}
This result is in essence similar to Lemma 4.3 of \cite{ozgur2}.
We only need to establish a few facts here. The minimum distance
between a pair of nodes, $i\in T$ and $j \in R$, is $\sqrt{g(K)}$,
and the maximum of such a distance equals $\sqrt{10 g(K)}$.
Furthermore, the distance between the center of transmit- and
receive-clusters equals $d_{TR}=2\sqrt{g(K)}$. Therefore, $d_{ij}$
follows $\frac{1}{2}d_{TR}\le d_{ij}\le
\frac{\sqrt{10}}{2}d_{TR}$, and hence,
$d^{\frac{-\alpha}{2}}_{ij}=d^{\frac{-\alpha}{2}}_{TR}\rho_{ij}$,
for $\rho_{ij} \in
[(\frac{2}{\sqrt{10}})^{\frac{\alpha}{2}},2^{\frac{\alpha}{2}}]$.
By treating the interference as a source of noise, and based on
our assumption that $P^{(2)}=2^\alpha P
g(K)^{\frac{\alpha}{2}-1}$, the mutual information between the two
clusters can be written as
\begin{equation}
\begin{split}
\nonumber &{\bf R_{\text{Inter}}}=I({\bf X};{\bf Y}, H)=
\log{\det(I+\frac{P^{(2)}}{N_0+\mathbb{I}_{\text{Inter}} } HH^{*})
}\ge
\\&\log{\det(I+\frac{  2^\alpha Pg(K)^{\frac{\alpha}{2}-1}
}{N_0+8 c_4 2^{\alpha}P} HH^{*}) }\\&= \log{\det(I+\frac{ P
}{c_5g(K) } FF^{*}) },
\end{split}
\end{equation}
\normalsize where ${\bf X}$, ${\bf Y}$, and $H$ are
respectively the corresponding input, output, and channel
matrices, $F$ is a $g(K)\times g(K)$ matrix with elements
$F_{ij}=\rho_{ij}e^{\theta_{ij}}$, and $c_5$ is a constant. The
rest of the problem has the same structure as Lemma 4.3 in
\cite{ozgur2} and the linearity of the rate growth with the
number of distributed antennas follows from there.

\section{Proof of Lemma 5}
The proof follows from Lemma 4.4 of \cite{ozgur2} and the fact that
the received power $P_r$ at any node $j$ in the receive-cluster
follows $P_r
=\sum_{i=1}^{g(K)}P^{(2)}|h_{ij}|^2\le\sum_{i=1}^{g(K)}P^{(2)}\frac{1}{g(K)^\frac{\alpha}{2}}=2^\alpha
Pg(K)^{\frac{\alpha}{2}-1}\sum_{i=1}^{g(K)}\frac{1}{g(K)^\frac{\alpha}{2}}=2^\alpha
P$, which is a constant.

\section{Proof of Lemma 6}
Assuming that there are $N$ nodes distributed uniformly over
$B_N$, for each node we define the random variable $S_i$ to be
equal to 1 if it lies in a given square with area $A\log{N}$
and 0 otherwise. Clearly, $P(S_i=1)=\frac{A\log{N}}{N}$. Using
Chernoff's bounds for binomial random variable
$S=\sum_{i=1}^{N}S_i$, which represents the number of nodes
inside a square, we have $p[S<(1-\delta)
N\frac{A\log{N}}{N}]=p[S<(1-\delta)A\log{N}]<e^{-A\log{N}\frac{\delta^2}{2}}=N^{-A\frac{\delta^2}{2}}$,
and, furthermore,
$p[S>(1+\delta)N\frac{A\log{N}}{N}]=p[S>(1+\delta)A\log{N}]<e^{-A\log{N}
f(\delta)}=N^{-Af(\delta)}$, where
$f(\delta)=(1+\delta)\log{(1+\delta)}$. To guarantee that the
number of nodes inside all the squares obeys these
inequalities, we use the union bound of the probabilities. It
directly follows that if
$A>\max(\frac{2}{\delta^2},\frac{1}{f(\delta)})$, with
probability 1 the number of nodes $M_j$ inside all squares of
area $A\log{N}$ follows
$(1-\delta)A\log{N}<M_j<(1+\delta)A\log{N}$, $ 1\le j\le N$ and
$\forall \delta>0$.

\end{appendices}

\begin{biography}[]{Sam Vakil}
received the B.Sc.~degree in Electrical Engineering from Sharif
University of Technology in Tehran, Iran, in 2002 and the M.Eng
degree in Electrical Engineering from McGill University in Montreal,
Canada, in 2004. He is now pursuing the Ph.D.~degree in Electrical \&
Computer Engineering at the University of Toronto. His current
research interest includes the design and analysis of cooperative
communication protocols for wireless ad-hoc and wireless networks.
\end{biography}

\begin{biography}[]{Ben Liang}
received honors simultaneous B.Sc.~(valedictorian) and M.Sc.~degrees
in electrical engineering from Polytechnic University in Brooklyn,
New York, in 1997 and the Ph.D. degree in electrical engineering
with computer science minor from Cornell University in Ithaca, New
York, in 2001. In the 2001 - 2002 academic year, he was a visiting
lecturer and post-doctoral research associate at Cornell University.
He joined the Department of Electrical and Computer Engineering at
the University of Toronto in 2002, where he is now an Associate
Professor. His current research interests are in mobile networking
and multimedia systems. He won an Intel Foundation Graduate
Fellowship in 2000 toward the completion of his Ph.D. dissertation
and an Early Researcher Award (ERA) given by the Ontario Ministry of
Research and Innovation in 2007. He was a co-author of the Best
Paper Award at the IFIP Networking conference in 2005 and the
Runner-up Best Paper Award at the International Conference on
Quality of Service in Heterogeneous Wired/Wireless Networks in 2006.
He is an editor for the IEEE Transactions on Wireless Communications
and an associate editor for the Wiley Security and Communication
Networks journal. He serves on the organizational and technical
committees of a number of conferences each year. He is a senior
member of IEEE and a member of ACM and Tau Beta Pi.
\end{biography}

\vfill

\end{document}